\documentclass[twocolumn,preprintnumbers,amsmath,amssymb,prb,superscriptaddress]{revtex4-2}
\usepackage{multirow}
\usepackage{amsfonts}
\usepackage[english]{babel}
\usepackage[T1]{fontenc}
\usepackage{times}
\usepackage{mathrsfs}
\usepackage{graphicx}
\usepackage{dcolumn}
\usepackage{bm}
\usepackage[colorlinks,bookmarks=true,citecolor=blue,linkcolor=red,urlcolor=blue]{hyperref}

\begin{document}
\title{Discovery of a dynamical quantum critical phase in a non-Hermitian system}
\author{Jiangrong Wen}
\affiliation{Department of Physics and Chongqing Key Laboratory for Strongly Coupled Physics, Chongqing University, Chongqing 401331, China}
\author{Zi-Xiang Hu}
\email[Contact author: ]{zxhu@cqu.edu.cn}
\affiliation{Department of Physics and Chongqing Key Laboratory for Strongly Coupled Physics, Chongqing University, Chongqing 401331, China}
\author{Jian-Jun Dong}
\email[Contact author: ]{dongjianjun@cqu.edu.cn}
\affiliation{Department of Physics and Chongqing Key Laboratory for Strongly Coupled Physics, Chongqing University, Chongqing 401331, China}
\date{\today}

\begin{abstract}
We identify dynamical critical phases in a non-Hermitian Kitaev chain with imbalanced pairing, revealing an extension from isolated critical points to critical phases in quantum quenches. By analyzing Fisher zeros and momentum-resolved Loschmidt return amplitudes, we show that continuous intervals of critical momenta yield zero times spanning continuous time intervals, with portions of the Fisher-zero branches lying on the imaginary axis of the complex-time plane. We derive the critical conditions for the four combinations of real and purely imaginary initial and final energies, establishing that the formation of dynamical critical phases is jointly governed by spectral types and state-overlap conditions. In particular, momentum intervals with purely imaginary initial energies and real final energies automatically satisfy the biorthogonal transition condition, whereas the self-normal formulation supports critical phases subject to additional constraints. These results demonstrate that dynamical critical phases are not exclusive to the biorthogonal description and provide a basis for exploring dynamical behavior beyond isolated critical points in non-Hermitian systems.
\end{abstract}

\maketitle

\section{Introduction}
\label{sec:introduction}

Dynamical quantum phase transitions (DQPTs) were first introduced by Heyl et al. in the quench dynamics of the transverse-field Ising model. Characterized by nonanalyticities of the Loschmidt rate function at critical times in the thermodynamic limit, DQPTs extend the concept of phase transitions to nonequilibrium quantum evolution \cite{Heyl2013}.
 Subsequent studies have explored various quench protocols \cite{Sharma2016,Bhattacharya2017Ramp,Dong2019}, mixed states \cite{Abeling2016,HeylBudich2017,Bhattacharya2017,
Sedlmayr2018,Lang2018Temperature}, and Floquet systems \cite{Yang2019Floquet,Zamani2020Floquet,Jafari2021Entanglement,ZhouDu2021Floquet,Cai2022Floquet,Naji2022Engineering,Jafari2022Synchronized}.
Experimental investigations have also been carried out in trapped ions \cite{Jurcevic2017}, ultracold atoms \cite{Flaschner2018}, superconducting qubits \cite{Guo2019Qubit},
and photonic quantum walks \cite{Wang2019Photonic,Xu2020DTOP}. In non-Hermitian systems, complex spectra and nonunitary evolution introduce new mechanisms for dynamical criticality
\cite{Zhou2018NH,ZhouDu2021NH,Mondal2022,Mondal2023,Naji2022Dissipative}.
Moreover, the biorthogonal structure of left and right eigenstates allows generally inequivalent definitions of state overlap
\cite{Leung1998,Brody2014,Ashida2020,Arkhipov2026,Jing2024Biorthogonal,Fu2025Anatomy,Zhang2025Metric}.
These definitions can yield different critical times
\cite{Jing2024Biorthogonal,Zhang2025Biorthogonal,Yao2026SSH,Gu2026Kitaev}
and even different predictions for whether a transition occurs
\cite{Wen2026Floquet}.
Non-Hermitian DQPTs thus provide a route to understanding dynamical criticality beyond the Hermitian framework and exploring new nonequilibrium critical phenomena.

Although non-Hermitian DQPTs exhibit critical-time distributions and topological features distinct from their Hermitian counterparts, the transitions in many existing quench studies still occur at isolated critical points in time
\cite{Zhou2018NH,ZhouDu2021NH,Mondal2022,Jing2024Biorthogonal,Zhang2025Biorthogonal}.
Equilibrium experiments, however, have revealed that quantum criticality can extend beyond isolated critical points to form a quantum-critical phase. In 2019, Zhao et al. found that magnetic frustration in the strongly correlated metal CePdAl stabilizes a quantum-critical phase over a finite region of parameter space
\cite{Zhao2019CriticalPhase}.
This finding motivates the search for analogous phenomena out of equilibrium \cite{Ramires2019Frustration}.
Meanwhile, the experimental discovery of an intrinsic non-Hermitian phase transition in a bulk material demonstrates that non-Hermitian dynamical criticality can arise in
solid-state systems \cite{Li2026IntrinsicNH}. These developments raise a natural question: Can DQPTs in non-Hermitian quantum quenches extend from isolated critical points to continuous intervals in time,thereby forming a dynamical critical phase?

Here, we address this question using a non-Hermitian Kitaev chain with imbalanced pairing. By analyzing Fisher zeros and momentum-resolved Loschmidt return amplitudes, we find that dynamical critical phases can emerge in both the biorthogonal and self-normalized formulations. In these phases, the zero times of the return amplitudes associated with continuous intervals of critical momenta cover continuous time intervals, with portions of the Fisher-zero branches lying on the imaginary axis $\operatorname{Re}z_n^X(k)=0$ of the complex-time plane. The formation of these phases is jointly determined by the initial and final spectral types and the state-overlap conditions. For momentum intervals with purely imaginary initial energies and real final energies, the biorthogonal transition condition is automatically satisfied, whereas the self-normalized formulation requires additional constraints. Consequently, the same quench can yield a dynamical critical phase in one formulation and isolated critical points in the other. These results reveal dynamical behavior beyond isolated critical points in non-Hermitian DQPTs and establish that dynamical critical phases are not exclusive to the biorthogonal description, providing a basis for exploring their generality in non-Hermitian systems.

\section{Theoretical framework}
\label{sec:theory}
For the initial and final non-Hermitian Hamiltonians $H_k^\alpha$, with $\alpha=i,f$, the right and left eigenstates satisfy
\begin{equation}
 H_k^\alpha|u_{k,s}^\alpha\rangle
 =E_{k,s}^\alpha|u_{k,s}^\alpha\rangle,\quad
 (H_k^\alpha)^\dagger|\widetilde u_{k,s}^\alpha\rangle
 =(E_{k,s}^\alpha)^*|\widetilde u_{k,s}^\alpha\rangle,
\end{equation}
where $s=\pm$ labels the two bands. At momenta where the Hamiltonian is diagonalizable, the eigenstates can be chosen to obey $ \langle\widetilde u_{k,s}^\alpha|u_{k,s'}^\alpha\rangle=\delta_{ss'}$. We exclude exceptional points of the initial Hamiltonian so that its biorthogonal eigenbasis is well defined. The initial state in each momentum sector is the right eigenstate $|u_{k,-}^i\rangle$ of the negative branch. After the quench, it evolves under the static Hamiltonian $H_k^f$. Setting $\hbar=1$, we write time-evolved state $ |u_k(t)\rangle=e^{-iH_k^ft}|u_{k,-}^i\rangle
 =\sum_{s=\pm}c_{k,s}(t)|u_{k,s}^i\rangle$
with $c_{k,s}(t)=\langle\widetilde u_{k,s}^i|u_k(t)\rangle$. The associated state is constructed using the same coefficients:
\begin{align}
 |\widetilde u_k(t)\rangle=\sum_{s=\pm}c_{k,s}(t)|\widetilde u_{k,s}^i\rangle,\quad
 \langle\widetilde u_k(t)|=\sum_{s=\pm}c_{k,s}^*(t)\langle\widetilde u_{k,s}^i|.
\end{align}
For the product initial state $|\Psi(0)\rangle=\bigotimes_k|u_{k,-}^i\rangle$, the biorthogonal Loschmidt echo is defined as~\cite{Jing2024Biorthogonal}
\begin{align}
 \mathcal L_B(t)=\frac{\langle\widetilde\Psi(0)|\Psi(t)\rangle
 \langle\widetilde\Psi(t)|\Psi(0)\rangle}
 {\langle\widetilde\Psi(0)|\Psi(0)\rangle
 \langle\widetilde\Psi(t)|\Psi(t)\rangle}
 =\prod_k g_k^B(t).
\end{align}
The momentum-resolved echo is
\begin{align}
 g_k^B(t)
=\frac{\langle\widetilde u_{k,-}^i|u_k(t)\rangle
 \langle\widetilde u_k(t)|u_{k,-}^i\rangle}
 {\langle\widetilde u_{k,-}^i|u_{k,-}^i\rangle
 \langle\widetilde u_k(t)|u_k(t)\rangle}=\frac{|c_{k,-}(t)|^2}{\sum_{s=\pm}|c_{k,s}(t)|^2}.
\end{align}
The corresponding normalized return amplitude reads
\begin{equation}
 \mathcal G_k^B(t)=
 \frac{\langle\widetilde u_{k,-}^i|u_k(t)\rangle}
 {\sqrt{\langle\widetilde u_{k,-}^i|u_{k,-}^i\rangle
 \langle\widetilde u_k(t)|u_k(t)\rangle}}.
\end{equation}
For comparison, the self-normal formulation uses the ordinary inner product:
\begin{equation}
 \mathcal G_k^S(t)=
 \frac{\langle u_{k,-}^i|u_k(t)\rangle}
 {\sqrt{\langle u_{k,-}^i|u_{k,-}^i\rangle
 \langle u_k(t)|u_k(t)\rangle}}.
\end{equation}
In either formulation, $g_k^X(t)=|\mathcal G_k^X(t)|^2$ and $\mathcal L_X(t)=\prod_k g_k^X(t)$, with $X=B,S$. The rate function is
\begin{equation}
 \lambda_X(t)=-\lim_{N\to\infty}\frac{1}{N}\ln\mathcal L_X(t),
 \qquad X=B,S,
\end{equation}
where $N$ is the real-space chain length. Zeros satisfying $\mathcal G_{k_c}^X(t_c)=0$ determine the critical zero set underlying possible nonanalytic structures of the thermodynamic rate function.

To analyze these zeros, we continue the return amplitude to complex time. We define $z=\tau+it$, so that physical evolution at $t>0$ corresponds to the positive imaginary axis ~\cite{Andraschko2014,Fisher1967,YangLee1952,LeeYang1952,ZaunerStauber2017}. For $H_k^f=\mathbf d_k^f\cdot\bm\sigma$, the identity $(H_k^f)^2=(E_k^f)^2I$ gives
\begin{equation}
 e^{-zH_k^f}=\cosh(E_k^fz)I
 -\frac{\sinh(E_k^fz)}{E_k^f}H_k^f,
\end{equation}
provided $E_k^f\ne0$. For $M_k^X\ne\pm1$, the Fisher zeros are
\begin{equation}
 z_n^X(k)=\frac{\operatorname{artanh} M_k^X}{E_k^f}
 +i\frac{(2n+1)\pi}{2E_k^f},\quad n\in\mathbb Z,
 \label{eq:fisher}
\end{equation}
where $ M_k^B=-\frac{\mathbf d_k^i\cdot\mathbf d_k^f}{E_k^iE_k^f}$ and $ M_k^S=\frac{\mathbf r_k^i\cdot\mathbf d_k^f}{E_k^f}$.
The dot product $\mathbf d_k^i\cdot\mathbf d_k^f$ does not involve complex conjugation. Here $\mathbf r_k^i=\langle u_{k,-}^i|\bm\sigma|u_{k,-}^i\rangle$, with the right initial state normalized under the ordinary inner product. Finite positive-time zeros require
\begin{equation}
 \operatorname{Re}z_n^X(k_c)=0,\qquad
 t_{c,n}^X=\operatorname{Im}z_n^X(k_c)>0.
 \label{eq:critical}
\end{equation}
In a non-Hermitian system, the two coefficients $M_k^X$ are generally inequivalent and can produce different Fisher-zero distributions.

\section{Dynamical critical points and phases in the non-Hermitian Kitaev chain}
\label{sec:results}

\subsection{Model and spectral structure}
\label{sec:model}
We consider a one-dimensional chain of spinless fermions with imbalanced $p$-wave pairing. Its real-space Hamiltonian is
\begin{align}
 \hat H={}&\sum_{j=1}^{N}\bigg[
 \mu\left(c_j^\dagger c_j-\frac12\right)
 +J\left(c_j^\dagger c_{j+1}+c_{j+1}^\dagger c_j\right)\bigg]\nonumber\\
 &+\frac12\sum_{j=1}^{N}\big[
 (\kappa-\Delta)c_j^\dagger c_{j+1}^\dagger
 -(\kappa+\Delta)c_{j+1}c_j\big].
 \label{eq:realH}
\end{align}
The operators $c_j^\dagger$ and $c_j$ create and annihilate a fermion at site $j$, respectively. The real parameters $J$, $\mu$, and $\Delta$ denote the nearest-neighbor hopping amplitude, chemical potential, and pairing parameter. The parameter $\kappa$ controls the imbalance between pair creation and annihilation. For $\kappa\ne0$, these pairing terms are no longer Hermitian conjugates of one another.

Under periodic boundary conditions, $c_{N+1}=c_1$, we use the Fourier transformation $c_j=N^{-1/2}\sum_k e^{ikj}c_k$ and introduce the Nambu spinor $\Psi_k^\dagger=(c_k^\dagger,c_{-k})$. The Hamiltonian separates into independent $(k,-k)$ sectors and it takes the form $\hat H=\sum_{0<k<\pi}\Psi_k^\dagger H_k\Psi_k$. We use an even chain length $N=2000$ for the momentum grid. The Bloch Hamiltonian is
\begin{equation}
 H_k=\begin{pmatrix}
 \xi_k&i(\kappa-\Delta)\sin k\\
 i(\kappa+\Delta)\sin k&-\xi_k
 \end{pmatrix}
 =\mathbf d_k\cdot\bm\sigma,
 \label{eq:bloch}
\end{equation}
where $\xi_k=\mu+2J\cos k$ and $\mathbf d_k=(i\kappa\sin k,\Delta\sin k,\xi_k)$. The eigenvalues are $E_{k,\pm}=\pm E_k$, with $ E_k=\sqrt{\xi_k^2+(\Delta^2-\kappa^2)\sin^2k}$. For real parameters, $\mathbf d_k^i\cdot\mathbf d_k^f$ is also real. We consider sudden quenches $(\mu_i,\kappa_i)\to(\mu_f,\kappa_f)$ at fixed $J$ and $\Delta$.

\begin{figure}[ptb]
\centering
\includegraphics[width=8.5cm]{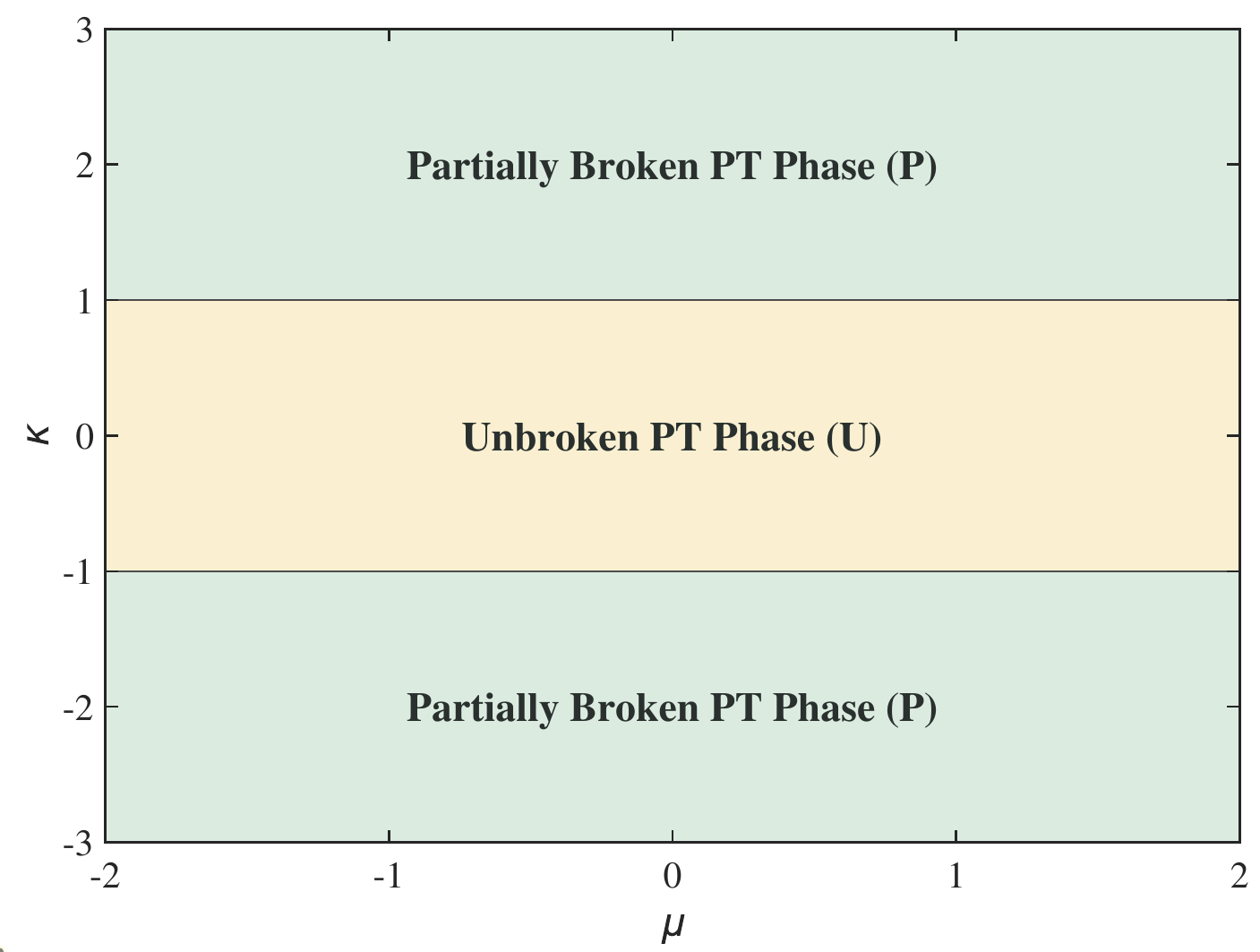}
\caption{Spectral phase diagram of the non-Hermitian Kitaev chain in the $(\mu,\kappa)$ plane for $J=\Delta=1$. The light beige region is the $\mathcal{PT}$-unbroken region (U), with real energies throughout the Brillouin zone. The light gray-green regions are partially $\mathcal{PT}$-broken (P), with real and purely imaginary energies in different momentum intervals.
}
\label{fig1}
\end{figure}

The model has $\mathcal{PT}$ symmetry and satisfies $E_k^2\in\mathbb R$. Positive and negative $E_k^2$ correspond to real and purely imaginary energies, respectively, whereas $E_k^2=0$ denotes a zero-energy gap closing. According to the spectrum over the Brillouin zone, the parameter range considered here is divided into a $\mathcal{PT}$-unbroken region (U) and a partially broken region (P), as shown in Fig.~\ref{fig1}. In U, the spectrum is real throughout the Brillouin zone. In P, real and purely imaginary energies occupy different momentum intervals joined at zero-energy degeneracies. Quenches are therefore classified as $U\to U$, $U\to P$, $P\to U$, or $P\to P$. Away from zero-energy degeneracies, each momentum belongs to one of four initial-to-final spectral combinations: real to real, real to purely imaginary, purely imaginary to real, or purely imaginary to purely imaginary. These momentum-resolved spectral combinations and the overlap conditions jointly determine the Fisher-zero distribution and its critical momentum and time structure.

\subsection{Critical conditions for different spectral combinations}
\label{sec:conditions}
\subsubsection*{i. Real to real}
If both initial and final energies are nonzero and real, $\mathbf d_k^i\cdot\mathbf d_k^f$ and $\mathbf r_k^i\cdot\mathbf d_k^f$ are real in this model. Thus, $M_k^X\in\mathbb R$ in both formulations. For $|M_k^X|<1$, Eq.~\eqref{eq:fisher} becomes
\begin{equation}
 z_n^X(k)=\frac{1}{2E_k^f}\ln\left|\frac{1+M_k^X}{1-M_k^X}\right|
 +i\frac{(2n+1)\pi}{2E_k^f}.
\end{equation}
The condition $\operatorname{Re}z_n^X(k)=0$ is equivalent to $M_k^X=0$. The critical momenta therefore satisfy
\begin{equation}
 \mathbf d_{k_c^B}^i\cdot\mathbf d_{k_c^B}^f=0,
 \qquad \mathbf r_{k_c^S}^i\cdot\mathbf d_{k_c^S}^f=0.
\end{equation}
For each such momentum, the critical times can be written  as
\begin{equation}
 t_{c,n}^X=\frac{(2n+1)\pi}{2E_{k_c^X}^f}.
 \label{eq:real_times}
\end{equation}
For finite real $|M_k^X|>1$, the Fisher zeros have nonzero real parts and do not intersect the imaginary axis of the complex $z$ plane. At $M_k^X=\pm1$, the overlap numerator reduces to the nonzero exponential $e^{\mp iE_k^ft}$ and has no finite-time zero. Hence, the condition throughout a real-to-real spectral interval is simply $M_k^X=0$. Generically, it selects isolated critical momenta, each associated with equally spaced zero times, as in Hermitian two-band quenches. This spectral sector therefore generically gives rise to isolated dynamical critical points. Although Eq.~\eqref{eq:real_times} has the same form for both formulations, their distinct critical momenta generally yield different critical times.
\subsubsection*{ii. Real to purely imaginary}
For nonzero real initial energies and nonzero purely imaginary final energies, both formulations satisfy $\operatorname{Re}M_k^X=0$ in this model. Using $\operatorname{artanh} M_k^X=i\arctan(\operatorname{Im}M_k^X)$, with branch changes absorbed into $n$, gives
\begin{equation}
 z_n^X(k)=\frac{(n+\tfrac12)\pi+\arctan(\operatorname{Im}M_k^X)}
 {\operatorname{Im}E_k^f},\quad n\in\mathbb Z.
\end{equation}
For finite $M_k^X$, the arctangent lies strictly between $-\pi/2$ and $\pi/2$. Consequently,
\begin{equation}
 \operatorname{Re}z_n^X(k)\ne0,\qquad\operatorname{Im}z_n^X(k)=0.
\end{equation}
These zeros lie on the real axis of the complex $z$ plane and do not intersect the imaginary axis $\operatorname{Re}z_n(k)=0$ . Neither formulation therefore supports dynamical critical points or a dynamical critical phase in this spectral sector.
\subsubsection*{iii. Purely imaginary to real}
For nonzero purely imaginary initial energies and nonzero real final energies, the biorthogonal coefficient automatically satisfies $\operatorname{Re}M_k^B=0$. The Fisher zeros are
\begin{equation}
 z_n^B(k)=i\frac{(n+\tfrac12)\pi+\arctan(\operatorname{Im}M_k^B)}
 {E_k^f},\quad n\in\mathbb Z,
\end{equation}
and the corresponding zero times are
\begin{equation}
 t_n^B(k)=\frac{(n+\tfrac12)\pi+\arctan(\operatorname{Im}M_k^B)}{E_k^f}.
 \label{eq:imag_real_times}
\end{equation}
The entire interval thus satisfies the biorthogonal real-time zero condition and forms a continuous set of critical momenta. Each momentum has a sequence of discrete zero times, which collectively define continuous branches $t_n^B(k)$. A continuous nonconstant positive-time branch covers a time interval, thereby forming a dynamical critical phase.

By contrast, $M_k^S$ is generally complex. The self-normal formulation has real-time zeros only at momenta satisfying $\operatorname{Re}M_k^S=0$. At those momenta, the Fisher zeros and critical times have the same functional form as above, with $B$ replaced by $S$. If this condition holds throughout a momentum interval, continuous critical momenta emerge and portions of the Fisher-zero branches lie on the imaginary axis $\operatorname{Re}z_n^S(k)=0$. If it holds only at isolated momenta, the critical momenta remain isolated. Thus, the biorthogonal formulation automatically satisfies the condition for the critical phase in this spectral sector, while the self-normal formulation retains an additional overlap constraint. This difference allows a single quench to display distinct critical structure in the two formulations.
\subsubsection*{iv.Purely imaginary to purely imaginary}
If both energies are nonzero and purely imaginary, $M_k^B$ is real in this model. A finite real-time zero requires $|M_k^B|>1$. We may then choose $ \operatorname{artanh} M_k^B=q_k^B-\frac{i\pi}{2}$ and $q_k^B=\frac12\ln\left|\frac{1+M_k^B}{1-M_k^B}\right|$, so that
\begin{equation}
 z_n^B(k)=\frac{n\pi}{\operatorname{Im}E_k^f}
 -i\frac{q_k^B}{\operatorname{Im}E_k^f},\quad n\in\mathbb Z.
\end{equation}
With this branch labeling, only $n=0$ satisfies $\operatorname{Re}z_n^B(k)=0$. Each qualifying momentum has a unique finite real-time zero,
\begin{equation}
 t_0^B(k)=-\frac{1}{2\operatorname{Im}E_k^f}
 \ln\left|\frac{1+M_k^B}{1-M_k^B}\right|.
 \label{eq:imag_imag_times}
\end{equation}
Positivity additionally requires $M_k^B\operatorname{Im}E_k^f<0$. Purely imaginary initial and final spectra alone therefore do not ensure positive-time zeros.

In the self-normal formulation, $M_k^S$ is generally complex, so the additional reality condition $\operatorname{Im}M_k^S=0$ must first be imposed. Together with $|M_k^S|>1$ and $M_k^S\operatorname{Im}E_k^f<0$, it gives the same Fisher-zero and critical-time expressions with $B$ replaced by $S$. In either formulation, each qualifying momentum contributes a unique positive-time zero through the $n=0$ branch. If these conditions hold throughout a continuous momentum interval, the interval forms a dynamical critical phase. A continuous nonconstant function $t_0^X(k)$ then covers a continuous time interval. The distinction is that the reality of $M_k^B$ is automatic in this spectral sector, whereas the reality of $M_k^S$ requires an additional overlap constraint; both formulations must still satisfy the magnitude and sign conditions.
\subsection{From dynamical critical points to critical phases}
\label{sec:examples}

\begin{figure}[ptb]
\centering
\includegraphics[width=8.5cm]{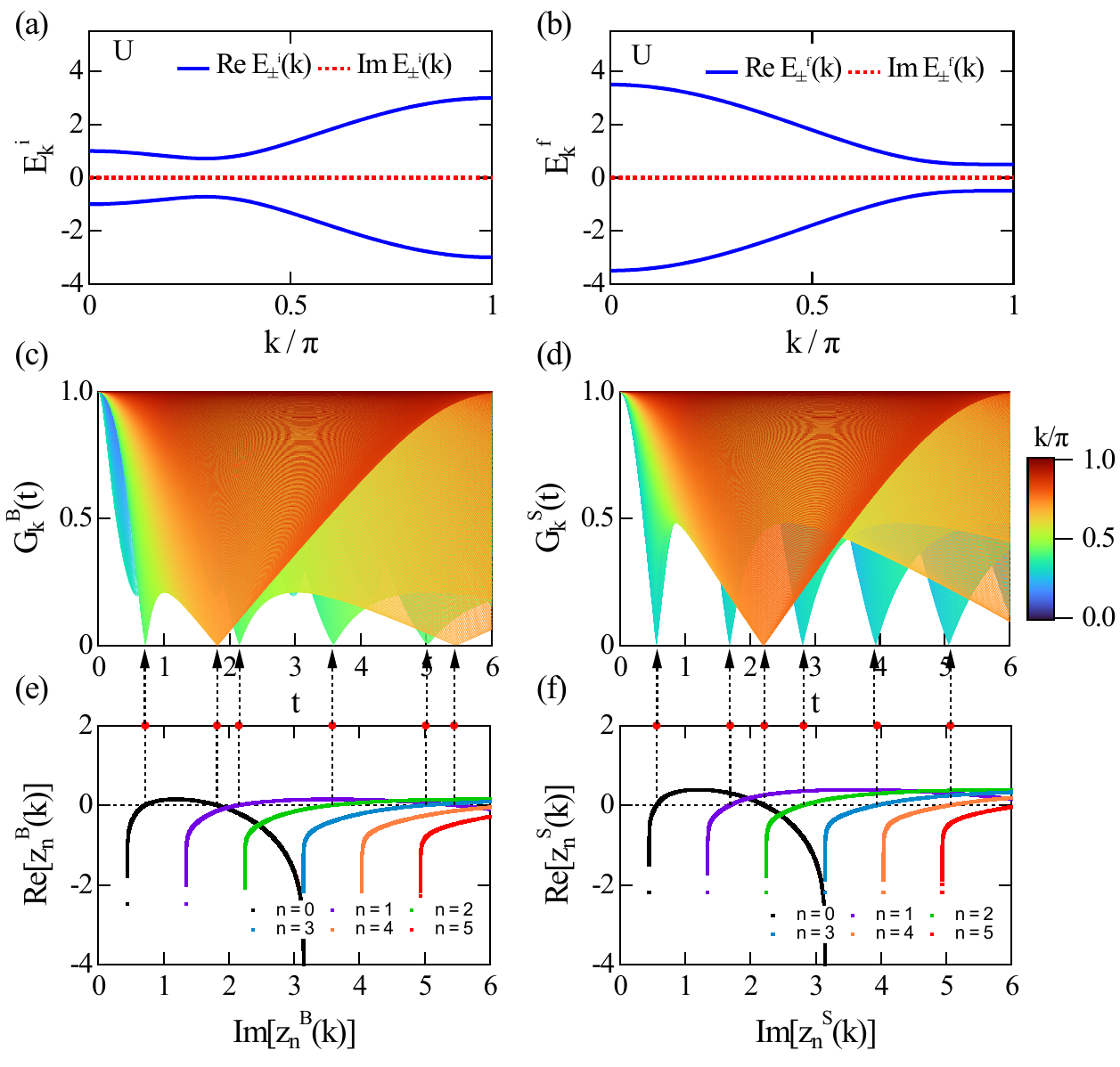}
\caption{Dynamical critical behavior in a $U\to U$ quench for $J=\Delta=1$, $(\mu_i,\kappa_i)=(-1,-0.5)$, and $(\mu_f,\kappa_f)=(1.5,0.2)$. (a),(b) Initial and final spectra; blue solid and red dashed curves denote the real and imaginary parts. (c),(d) Moduli of the biorthogonal and self-normal return amplitudes, colored by $k/\pi$. (e),(f) Corresponding Fisher zeros, with colors labeling different branches. The horizontal dashed line denotes the imaginary axis $\operatorname{Re}z_n(k)=0$. Arrows, red dots, and vertical dashed lines mark critical times and connect the two lower rows. Both formulations exhibit isolated critical points, but their critical times differ.
}
\label{fig2}
\end{figure}

We now examine representative $U\to U$, $P\to P$, and $P\to U$ quenches to illustrate isolated dynamical critical points, dynamical critical phases in both formulations,
and their formulation-dependent occurrence, respectively. The results refer to the selected parameters rather than to all quenches with the same U/P labels.

For the $U\to U$ quench, both Hamiltonians have real spectra throughout the Brillouin zone in Figs.~\ref{fig2} (a) and (b). At the selected parameters, both formulations exhibit isolated dynamical critical points arising from isolated critical momenta. The biorthogonal critical momenta are $k_{c,1}^B=0.42382070\pi$ and $k_{c,2}^B=0.69768027\pi$, whereas the self-normal ones are $k_{c,1}^S=0.29947943\pi$ and $k_{c,2}^S=0.74542235\pi$. At each critical momentum, the return amplitude vanishes periodically, producing discrete critical times in Figs.~\ref{fig2} (c) and (d). The Fisher-zero branches intersect the imaginary axis at the same times in Figs.~\ref{fig2} (e) and (f). Although the critical-time expression is common to both formulations, their different critical momenta yield different time distributions. This real-spectrum quench retains the isolated-zero structure familiar from Hermitian two-band systems, while the non-Hermitian overlap structure shifts the critical points.

\begin{figure}[ptb]
\centering
\includegraphics[width=8.5cm]{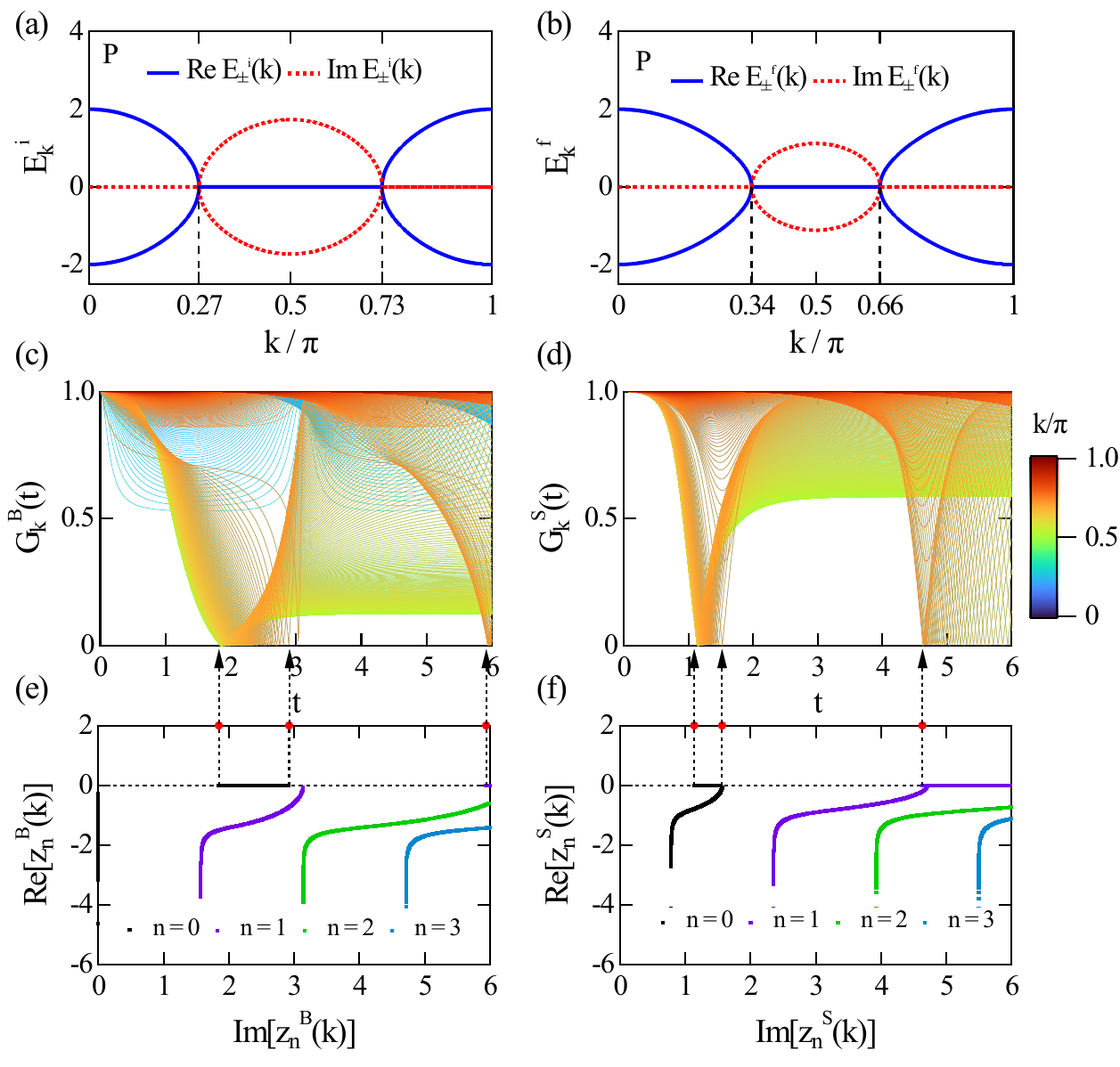}
\caption{Continuous critical structures in a $P\to P$ quench for $J=\Delta=1$, $(\mu_i,\kappa_i)=(0,-2)$, and $(\mu_f,\kappa_f)=(0,-1.5)$. (a),(b) Initial and final spectra, with vertical dashed lines marking gap-closing momenta. (c),(d) Moduli of the biorthogonal and self-normal return amplitudes, colored by $k/\pi$. (e),(f) Corresponding Fisher-zero branches. Arrows, red dots, and vertical dashed lines identify selected boundaries of zero-time intervals and their correspondence between rows. Both formulations exhibit dynamical critical phases, but the corresponding time intervals differ.
}
\label{fig3}
\end{figure}

For the $P\to P$ quench, both spectra contain real and purely imaginary momentum intervals in Figs.~\ref{fig3} (a) and (b). Excluding the gap-closing points, the selected quench contains real-to-real, purely imaginary-to-real, and purely imaginary-to-purely imaginary sectors. Neither formulation has a critical momentum in the real-to-real sector. In the purely imaginary-to-real sector $k\in{}(0.272817\pi,0.337750\pi)\nonumber\\
\cup(0.662250\pi,0.727183\pi),$ both formulations satisfy $\operatorname{Re}M_k^X=0$, producing continuous critical momenta. Each momentum contributes a sequence of discrete critical times that vary continuously with $k$. The return amplitudes vanish at their respective times, and portions of different Fisher-zero branches lie on the imaginary axis in Figs.~\ref{fig3} (c)--(f).

In the purely imaginary-to-purely imaginary sector, $k\in(0.337750\pi,0.662250\pi)$, both formulations satisfy $M_k^X\in\mathbb R$, $|M_k^X|>1$, and $M_k^X\operatorname{Im}E_k^f<0$ for the selected parameters. Each momentum therefore has a unique positive critical time $t_0^X(k)$. With the branch convention above, only the $n=0$ Fisher-zero branch lies on the imaginary axis in this sector; all other branches remain off the axis in Figs.~\ref{fig3} (e) and (f). At these parameters, both formulations exhibit dynamical critical phases. Although the two formulations share the same critical momentum intervals in this example, their different coefficients $M_k^X$ lead to dynamical critical phases covering different time intervals.

\begin{figure}[ptb]
\centering
\includegraphics[width=8.5cm]{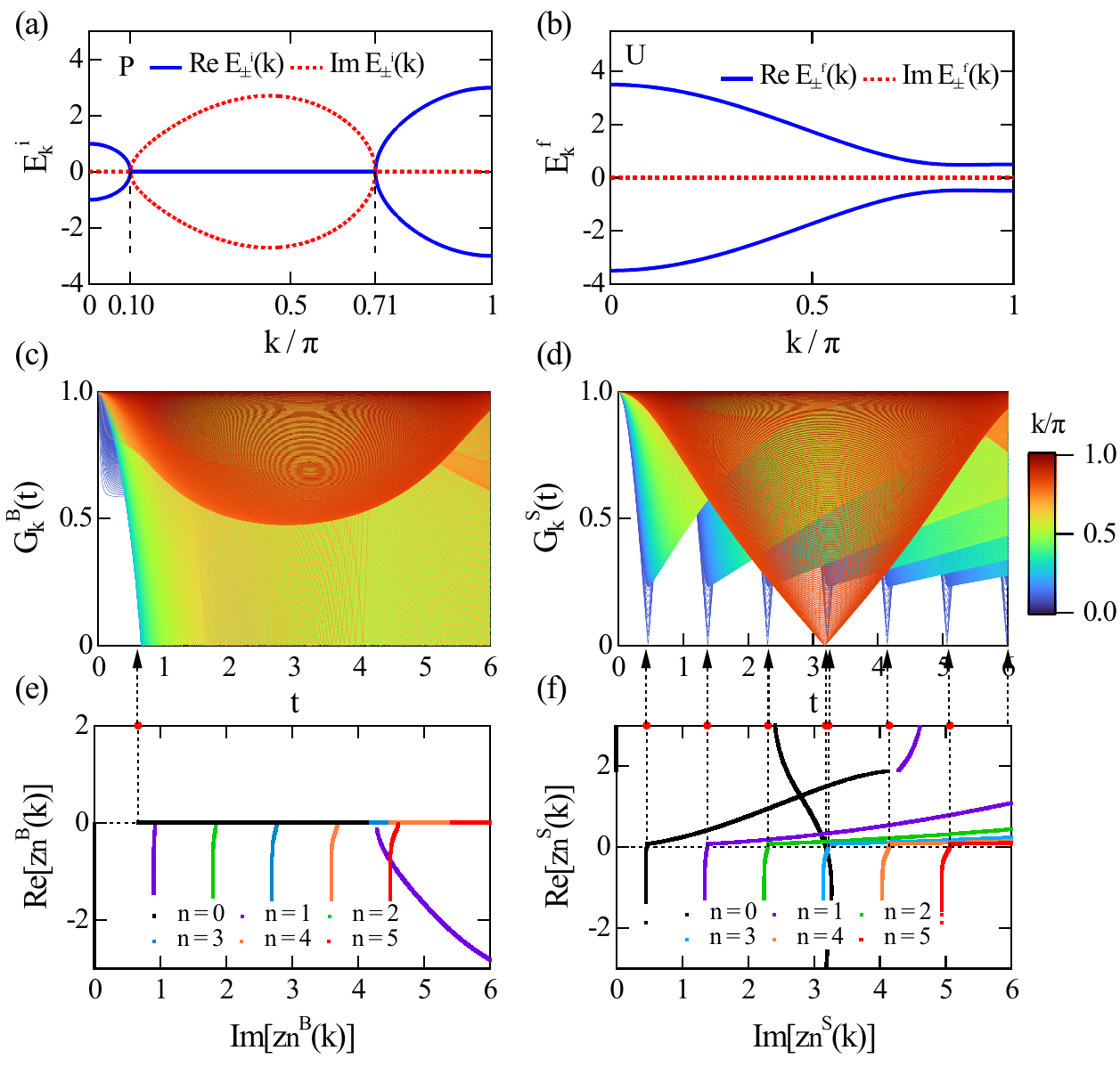}
\caption{Formulation-dependent continuous and isolated critical structures in a $P\to U$ quench for $J=\Delta=1$, $(\mu_i,\kappa_i)=(-1,-3)$, and $(\mu_f,\kappa_f)=(1.5,0.5)$. (a),(b) Initial and final spectra; vertical dashed lines mark the initial gap-closing momenta. (c),(d) Moduli of the biorthogonal and self-normal return amplitudes, colored by $k/\pi$. (e),(f) Corresponding Fisher-zero branches. Arrows, red dots, and vertical dashed lines indicate the onset of the displayed continuous zero-time interval or isolated critical times. The biorthogonal formulation exhibits a dynamical critical phase, whereas the self-normal formulation exhibits isolated critical points.}
\label{fig4}
\end{figure}

For this $P\to U$ quench at the selected parameters, the two formulations do not show identical critical structures; the biorthogonal formulation features a dynamical critical phase, while the self-normal formulation yields isolated critical points. The initial spectrum contains real and purely imaginary intervals joined at zero-energy degeneracies, whereas the final spectrum is real throughout the Brillouin zone in Figs.~\ref{fig4} (a) and (b). Away from the gap closings, the quench divides into real-to-real and purely imaginary-to-real sectors. In the real-to-real sector, the biorthogonal formulation has no critical momentum, while the self-normal formulation has two isolated critical momenta, $k_{c,1}^S=0.099940\pi$ and $k_{c,2}^S=0.828915\pi$. Their return amplitudes vanish periodically in Fig.~\ref{fig4} (d), and the self-normal Fisher-zero branches intersect the imaginary axis at isolated times in Fig.~\ref{fig4} (f).

In the purely imaginary-to-real interval, $k\in(0.102698\pi,0.710854\pi)$, the biorthogonal condition $\operatorname{Re}M_k^B=0$ holds automatically. The corresponding critical times vary continuously with momentum in Fig.~\ref{fig4} (c), and portions of the biorthogonal Fisher-zero branches lie on the imaginary axis in Fig.~\ref{fig4} (e). The self-normal coefficient does not satisfy $\operatorname{Re}M_k^S=0$ in this interval for the selected parameters, so it has no corresponding critical momenta. The same quench thus yields a dynamical critical phase in the biorthogonal formulation and isolated critical points in the self-normal formulation. The choice of state overlap can therefore distinguish critical phases from critical points, rather than merely shift the critical times.

\section{Conclusion}
\label{sec:conclusion}

We have investigated dynamical critical points and phases in quantum quenches of a $\mathcal{PT}$-symmetric non-Hermitian Kitaev chain. Combining analytical Fisher-zero
conditions with momentum-resolved Loschmidt return amplitudes, we find that both the biorthogonal and self-normal formulations can support dynamical critical phases. Their formation is jointly governed by the initial and final spectral types and the state-overlap conditions. Consequently, the two formulations can differ not only in their critical times but also in whether they exhibit isolated critical points or dynamical critical phases.

Away from zero-energy degeneracies, the four spectral combinations impose distinct zero conditions. For real initial and final energies, both formulations require $M_k^X=0$, which generically selects isolated critical momenta and discrete sequences of critical times. Real initial and purely imaginary final energies admit no finite positive-time zeros. Purely imaginary initial and real final energies automatically satisfy $\operatorname{Re}M_k^B=0$, whereas the self-normal formulation requires the additional condition $\operatorname{Re}M_k^S=0$. When both energies are purely imaginary, the conditions are $M_k^X\in\mathbb R$, $|M_k^X|>1$, and $M_k^X\operatorname{Im}E_k^f<0$; each qualifying momentum then has a unique positive critical time. In the latter two cases, conditions satisfied throughout a momentum interval yield continuous critical momenta. Whenever the associated time branch varies continuously with momentum, it covers a continuous time interval. Thus, the global U/P classification alone does not determine the critical structure; the momentum-resolved spectral combinations and overlap constraints are also essential.

In particular, a momentum interval with purely imaginary initial energies and real final energies automatically satisfies the biorthogonal transition condition, whereas
the self-normal formulation requires an additional constraint. The same quench can therefore exhibit a dynamical critical phase in one formulation and isolated critical points in the other. These findings reveal dynamical behavior beyond isolated critical points in non-Hermitian DQPTs. The occurrence of dynamical critical phases in both formulations further shows that they are not exclusive to the biorthogonal description, providing a basis for exploring their generality in non-Hermitian systems.


This work is supported by the National Natural Science Foundation of China Grants No. 12204075 and No. 12547101, the Natural Science Foundation of Chongqing Grant No. CSTB2023NSCQ-MSX0953 and No. CSTB2024YCJH-KYXM0064.
Z.~H was supported by the National Natural Science Foundation of China Grant No.12474140 and the Fundamental Research Funds for the Central Universities Grant No. 2025CDJ-IAISYB-029.


\begin{thebibliography}{99}
\bibitem{Heyl2013}
M. Heyl, A. Polkovnikov, and S. Kehrein,
Dynamical quantum phase transitions in the transverse-field Ising model,
Phys. Rev. Lett. \textbf{110}, 135704 (2013).

\bibitem{Sharma2016}
S. Sharma, U. Divakaran, A. Polkovnikov, and A. Dutta,
Slow quenches in a quantum Ising chain: Dynamical phase transitions and topology,
Phys. Rev. B \textbf{93}, 144306 (2016).

\bibitem{Bhattacharya2017Ramp}
U. Bhattacharya and A. Dutta,
Interconnections between equilibrium topology and dynamical quantum phase
transitions in a linearly ramped Haldane model,
Phys. Rev. B \textbf{95}, 184307 (2017).

\bibitem{Dong2019}
J.-J. Dong and Y.-F. Yang,
Functional field integral approach to quantum work,
Phys. Rev. B \textbf{100}, 035124 (2019).

\bibitem{Abeling2016}
N. O. Abeling and S. Kehrein,
Quantum quench dynamics in the transverse field Ising model at nonzero temperatures,
Phys. Rev. B \textbf{93}, 104302 (2016).

\bibitem{HeylBudich2017}
M. Heyl and J. C. Budich,
Dynamical topological quantum phase transitions for mixed states,
Phys. Rev. B \textbf{96}, 180304(R) (2017).

\bibitem{Bhattacharya2017}
U. Bhattacharya, S. Bandyopadhyay, and A. Dutta,
Mixed state dynamical quantum phase transitions,
Phys. Rev. B \textbf{96}, 180303(R) (2017).

\bibitem{Sedlmayr2018}
N. Sedlmayr, M. Fleischhauer, and J. Sirker,
Fate of dynamical phase transitions at finite temperatures and in open systems,
Phys. Rev. B \textbf{97}, 045147 (2018).

\bibitem{Lang2018Temperature}
J. Lang, B. Frank, and J. C. Halimeh,
Concurrence of dynamical phase transitions at finite temperature
in the fully connected transverse-field Ising model,
Phys. Rev. B \textbf{97}, 174401 (2018).

\bibitem{Yang2019Floquet}
K. Yang, L. Zhou, W. Ma, X. Kong, P. Wang, X. Qin,
X. Rong, Y. Wang, F. Shi, J. Gong, and J. Du,
Floquet dynamical quantum phase transitions,
Phys. Rev. B \textbf{100}, 085308 (2019).

\bibitem{Zamani2020Floquet}
S. Zamani, R. Jafari, and A. Langari,
Floquet dynamical quantum phase transition in the extended XY model:
Nonadiabatic to adiabatic topological transition,
Phys. Rev. B \textbf{102}, 144306 (2020).

\bibitem{Jafari2021Entanglement}
R. Jafari and A. Akbari,
Floquet dynamical phase transition and entanglement spectrum,
Phys. Rev. A \textbf{103}, 012204 (2021).

\bibitem{ZhouDu2021Floquet}
L. Zhou and Q. Du,
Floquet dynamical quantum phase transitions in periodically quenched systems,
J. Phys.: Condens. Matter \textbf{33}, 345403 (2021).

\bibitem{Cai2022Floquet}
D.-H. Cai and W. Yi,
Synthetic topology and Floquet dynamic quantum phase transition
in a periodically driven Raman lattice,
Phys. Rev. A \textbf{105}, 042812 (2022).

\bibitem{Naji2022Engineering}
J. Naji, R. Jafari, L. Zhou, and A. Langari,
Engineering Floquet dynamical quantum phase transitions,
Phys. Rev. B \textbf{106}, 094314 (2022).

\bibitem{Jafari2022Synchronized}
R. Jafari, A. Akbari, U. Mishra, and H. Johannesson,
Floquet dynamical quantum phase transitions under synchronized periodic driving,
Phys. Rev. B \textbf{105}, 094311 (2022).

\bibitem{Jurcevic2017}
P. Jurcevic, H. Shen, P. Hauke, C. Maier, T. Brydges,
C. Hempel, B. P. Lanyon, M. Heyl, R. Blatt, and C. F. Roos,
Direct observation of dynamical quantum phase transitions
in an interacting many-body system,
Phys. Rev. Lett. \textbf{119}, 080501 (2017).

\bibitem{Flaschner2018}
N. Fl{\"a}schner, D. Vogel, M. Tarnowski, B. S. Rem,
D.-S. L{\"u}hmann, M. Heyl, J. C. Budich, L. Mathey,
K. Sengstock, and C. Weitenberg,
Observation of dynamical vortices after quenches in a system with topology,
Nat. Phys. \textbf{14}, 265 (2018).

\bibitem{Guo2019Qubit}
X.-Y. Guo, C. Yang, Y. Zeng, Y. Peng, H.-K. Li,
H. Deng, Y.-R. Jin, S. Chen, D. Zheng, and H. Fan,
Observation of a dynamical quantum phase transition
by a superconducting qubit simulation,
Phys. Rev. Appl. \textbf{11}, 044080 (2019).

\bibitem{Wang2019Photonic}
K. Wang, X. Qiu, L. Xiao, X. Zhan, Z. Bian, W. Yi, and P. Xue,
Simulating dynamic quantum phase transitions in photonic quantum walks,
Phys. Rev. Lett. \textbf{122}, 020501 (2019).

\bibitem{Xu2020DTOP}
X.-Y. Xu, Q.-Q. Wang, M. Heyl, J. C. Budich,
W.-W. Pan, Z. Chen, M. Jan, K. Sun, J.-S. Xu,
Y.-J. Han, C.-F. Li, and G.-C. Guo,
Measuring a dynamical topological order parameter in quantum walks,
Light Sci. Appl. \textbf{9}, 7 (2020).

\bibitem{Zhou2018NH}
L. Zhou, Q.-H. Wang, H. Wang, and J. Gong,
Dynamical quantum phase transitions in non-Hermitian lattices,
Phys. Rev. A \textbf{98}, 022129 (2018).

\bibitem{ZhouDu2021NH}
L. Zhou and Q. Du,
Non-Hermitian topological phases and dynamical quantum phase transitions:
A generic connection,
New J. Phys. \textbf{23}, 063041 (2021).

\bibitem{Mondal2022}
D. Mondal and T. Nag,
Anomaly in the dynamical quantum phase transition
in a non-Hermitian system with extended gapless phases,
Phys. Rev. B \textbf{106}, 054308 (2022).

\bibitem{Mondal2023}
D. Mondal and T. Nag,
Finite-temperature dynamical quantum phase transition in a non-Hermitian system,
Phys. Rev. B \textbf{107}, 184311 (2023).

\bibitem{Naji2022Dissipative}
J. Naji, M. Jafari, R. Jafari, and A. Akbari,
Dissipative Floquet dynamical quantum phase transition,
Phys. Rev. A \textbf{105}, 022220 (2022).

\bibitem{Leung1998}
P. T. Leung, W. M. Suen, C. P. Sun, and K. Young,
Waves in open systems via a biorthogonal basis,
Phys. Rev. E \textbf{57}, 6101 (1998).

\bibitem{Brody2014}
D. C. Brody,
Biorthogonal quantum mechanics,
J. Phys. A: Math. Theor. \textbf{47}, 035305 (2014).

\bibitem{Ashida2020}
Y. Ashida, Z. Gong, and M. Ueda,
Non-Hermitian physics,
Adv. Phys. \textbf{69}, 249 (2020).

\bibitem{Arkhipov2026}
I. I. Arkhipov,
Covariant formulation of the Berry connection in non-Hermitian systems,
Phys. Rev. Res. \textbf{8}, 033037 (2026).

\bibitem{Jing2024Biorthogonal}
Y. Jing, J.-J. Dong, Y.-Y. Zhang, and Z.-X. Hu,
Biorthogonal dynamical quantum phase transitions in non-Hermitian systems,
Phys. Rev. Lett. \textbf{132}, 220402 (2024).

\bibitem{Fu2025Anatomy}
Y. Fu and G. Xianlong,
Anatomy of non-Hermitian dynamical quantum phase transitions,
Phys. Rev. B \textbf{112}, L140302 (2025).

\bibitem{Zhang2025Metric}
X. Zhang, K. Hu, and F. Li,
Nonadiabatic dynamics in non-Hermitian quantum systems
from the quantum metric perspective,
Phys. Rev. B \textbf{112}, 224305 (2025).

\bibitem{Zhang2025Biorthogonal}
H. Zhang, K. Wang, L. Xiao, and P. Xue,
Self-normal and biorthogonal dynamical quantum phase transitions
in non-Hermitian quantum walks,
Light Sci. Appl. \textbf{14}, 253 (2025).

\bibitem{Yao2026SSH}
J. Yao, R. Zhang, S. Cheng, Z. Wang, and H. Gu,
Dynamical quantum phase transition
in the non-Hermitian generalized Su--Schrieffer--Heeger model,
Eur. Phys. J. B \textbf{99}, 69 (2026).

\bibitem{Gu2026Kitaev}
H. Gu, Y. Zhao, S. Cheng, Y. Xie, X. Yang, and Y. Chen,
Biorthogonal dynamical quantum phase transitions
in non-Hermitian topological superconductors,
arXiv:2605.15982v2 (2026).

\bibitem{Wen2026Floquet}
J. Wen, Q. Yuan, Z.-X. Hu, and J.-J. Dong,
Biorthogonal-only Floquet dynamical quantum phase transitions,
arXiv:2608.07882v1 (2026).

\bibitem{Zhao2019CriticalPhase}
H. Zhao, J. Zhang, M. Lyu, S. Bachus, Y. Tokiwa,
P. Gegenwart, S. Zhang, J. Cheng, Y.-F. Yang,
G. Chen, Y. Isikawa, Q. Si, F. Steglich, and P. Sun,
Quantum-critical phase from frustrated magnetism
in a strongly correlated metal,
Nat. Phys. \textbf{15}, 1261 (2019).

\bibitem{Ramires2019Frustration}
A. Ramires,
Frustration can be critical,
Nat. Phys. \textbf{15}, 1212 (2019).

\bibitem{Li2026IntrinsicNH}
J. Li, M. Turaev, M. Matsubara, K. Kliemt,
C. Krellner, S. Pal, M. Fiebig, and J. Kroha,
Discovery of an intrinsic non-Hermitian phase transition
in a bulk condensed-matter system,
Science \textbf{393}, 1152 (2026).


\bibitem{Andraschko2014}
F. Andraschko and J. Sirker,
Dynamical quantum phase transitions and the Loschmidt echo:
A transfer matrix approach,
Phys. Rev. B \textbf{89}, 125120 (2014).


\bibitem{Fisher1967}M. E. Fisher,
The theory of equilibrium critical phenomena,
Rep. Prog. Phys. \textbf{30}, 615 (1967).

\bibitem{YangLee1952}C. N. Yang and T. D. Lee,
Statistical theory of equations of state and phase transitions.
I. Theory of condensation,
Phys. Rev. \textbf{87}, 404 (1952).

\bibitem{LeeYang1952}T. D. Lee and C. N. Yang,
Statistical theory of equations of state and phase transitions.
II. Lattice gas and Ising model,
Phys. Rev. \textbf{87}, 410 (1952).

\bibitem{ZaunerStauber2017}V. Zauner-Stauber and J. C. Halimeh,
Probing the anomalous dynamical phase in long-range quantum
spin chains through Fisher-zero lines,
Phys. Rev. E \textbf{96}, 062118 (2017).




\end{thebibliography}
\end{document}